\documentclass[pra,twocolumn,aps,amssymb,footinbib]{revtex4}

\usepackage{amssymb}
\usepackage{graphicx}
\usepackage{amsmath}
\usepackage{times}
\usepackage{color}
\usepackage{subfigure}
\usepackage{setspace}
\usepackage{bm}
\usepackage{xr}

\begin{document}




\date{\today}

\begin{abstract}
{\bf Fermionic alkaline-earth atoms have unique properties that make them attractive candidates for the realization of novel atomic clocks and degenerate quantum gases. At the same time,  they are attracting
considerable theoretical attention in the context of quantum information processing. Here we demonstrate that when such atoms are loaded in optical lattices, they can be used as  quantum simulators of unique many-body phenomena. In particular, we show  that the decoupling of the nuclear spin from the electronic angular momentum can be used to implement many-body  systems with an unprecedented degree of symmetry, characterized by the SU(N) group with N as large as 10. Moreover,  the interplay of the nuclear spin with the electronic degree of freedom provided  by a stable optically excited state allows for the study of spin-orbital physics. Such systems may provide valuable  insights into strongly correlated physics of transition metal oxides, heavy fermion materials, and spin liquid phases.}
\end{abstract}

\title{Two-orbital SU(N) magnetism with ultracold alkaline-earth atoms}

\author{A. V. Gorshkov$^{1*}$,  M. Hermele$^{2}$, V. Gurarie$^{2}$,  C. Xu$^{1}$, P. S. Julienne$^{3}$, J. Ye$^{4}$, P. Zoller$^{5}$, E. Demler$^{1,6}$, M. D.
Lukin$^{1,6}$,  and A. M. Rey$^{4}$}

\affiliation{$^{1}$Physics Department, Harvard University, Cambridge, MA 02138}
\affiliation{$^{2}$Department of Physics, University of Colorado, Boulder, CO 80309}
\affiliation{$^{3}$Joint Quantum Institute, NIST and University of Maryland, Gaithersburg, MD 20899-8423}
\affiliation{$^{4}$JILA, NIST, and Department of Physics, University of Colorado, Boulder, CO 80309}
\affiliation{$^{5}$Institute for Theoretical Physics, University of Innsbruck, A-6020 Innsbruck, Austria
and Institute for Quantum Optics and Quantum Information of the Austrian Academy of Sciences, A-6020 Innsbruck, Austria}
\affiliation{$^{6}$Institute for Theoretical Atomic, Molecular and Optical Physics,
Harvard-Smithsonian Center of Astrophysics, Cambridge, MA 02138}
\affiliation{$^{*}$e-mail: gorshkov@post.harvard.edu}
\maketitle


The interest in fermionic alkaline-earth atoms \cite{Boyd2007,Campbell2009,Lemke2009,Fukuhara2007b,Reichenbach2007, Hayes2007, Daley2008, Gorshkov2009}  stems from their two key features:  (1) the presence
of a metastable excited state ${}^3P_0$ coupled to the ground
${}^1S_0$ state via an ultranarrow doubly-forbidden transition
\cite{Boyd2007} and (2) the almost perfect decoupling \cite{Boyd2007} 
of the nuclear spin $I$ from the electronic
angular momentum $J$ in these two states, since they both have $J =
0$. This decoupling implies
that s-wave scattering lengths involving states ${}^1S_0$ and
${}^3P_0$ are independent of the nuclear spin, aside from the
restrictions imposed by fermionic antisymmetry. We  show that the resulting SU(N) spin symmetry (where $N = 2 I
+ 1$ can be as large as 10) together with the possibility of
combining (nuclear) spin physics with (electronic) orbital physics
open up a wide field of extremely rich many-body systems with
alkaline-earth atoms.

In what follows, we derive the two-orbital SU(N)-symmetric Hubbard model describing alkaline-earth atoms in  ${}^1S_0$ and ${}^3P_0$ states trapped in an optical lattice. We focus on specific parameter regimes characterized by full or partial atom localization due to strong atomic interactions, 
where simpler effective spin Hamiltonians can be derived. The interplay between  orbital and spin  degrees of freedom in such effective models is a central topic in quantum magnetism and has attracted  tremendous interest in the condensed matter community. Alkaline earth atoms thus provide, on the one hand,  a unique opportunity for the implementation of some of these models  for the first time in a defect-free and fully controllable  
environment.  On the other hand, they open a new arena to study a wide range of models,
many of which have not been discussed previously, even theoretically.
We demonstrate, in particular, how to implement  the Kugel-Khomskii model
studied in the context of transition metal oxides
\cite{Kugel1973b, Tokura2000,Li1998,Arovas1995,Pati1998},
the Kondo lattice model \cite{Ruderman1954,Coqblin1969,Doniach1977,Coleman1983,Tsunetsugu1997,Assaad1999,Tokura2000book,Oshikawa2000,Senthil2003,Duan2004,Paredes2005,Coleman2007,Gegenwart2008}
studied in context of manganese oxide
perovskites \cite{Tokura2000book} and heavy fermion materials \cite{Coleman2007}, 
as well as various
SU(N)-symmetric spin Hamiltonians that are believed to have spin
liquid and valence-bond-solid ground states \cite{Read1989b,Marston1989,Greiter2007,Xu2008,Harada2003,Assaad2005,Paramekanti2007,Hermele2009}.
%
For example, we
discuss how,  by appropriately choosing the initial state,  
a single alkaline-earth atom species with $I =
9/2$ (such as $^{87}$Sr) can be used to study experimentally such
a distinctively theoretical object as the phase diagram as a
function of $N$ for all $N \leq 10$.


Before proceeding, we note that, while an
orthogonal symmetry group SO(5) can be realized in alkali atoms \cite{Wu2003},  proposals to obtain SU(N$>$2)-symmetric
models with alkali atoms \cite{Honerkamp2004,Rapp2008} and solid state
systems \cite{Affleck1991,Li1998} are a substantial idealization due to strong
hyperfine coupling and a complex solid state environment, 
respectively. 
In this context, alkaline-earth-like atoms make a truly exceptional system to study
models with SU(N$>$2) symmetry.

\newpage

\section*{Many-body dynamics of alkaline-earth atoms in an optical lattice}

We begin with the Hamiltonian describing cold fermionic alkaline-earth atoms in an external trapping potential:
\begin{eqnarray}
\!\!\!\!\!&& H  =  \sum_{\alpha m} \int d^3 \mathbf{r} \Psi^\dagger_{\alpha m} (\mathbf{r}) (- \frac{\hbar^2}{2 M} \nabla^2 + V_{\alpha}(\mathbf{r})) \Psi_{\alpha m}(\mathbf{r})   \label{ham0} \\
\!\!\!\!\!&& + \hbar \omega_0\int d^3 \mathbf{r} (\rho_e(\mathbf{r})- \rho_g(\mathbf{r}))  + \frac{g_{eg}^+ + g_{eg}^-}{2}
\int d^3 \mathbf{r} \rho_e(\mathbf{r}) \rho_g(\mathbf{r})  \nonumber \\
\!\!\!\!\!&&   +\sum_{\alpha, m < m'} g_{\alpha \alpha}
\int d^3\mathbf{r} \rho_{\alpha m}(\mathbf{r}) \rho_{\alpha m'}(\mathbf{r})     \nonumber \\
\!\!\!\!\!&& + \frac{g_{eg}^+ - g_{eg}^-}{2}
\sum_{m m'} \int d^3 \mathbf{r} \Psi^\dagger_{g m}(\mathbf{r}) \Psi^\dagger_{e m'}(\mathbf{r}) \Psi_{g m'}(\mathbf{r}) \Psi_{e m}(\mathbf{r}). \nonumber
\end{eqnarray}
Here $\Psi_{\alpha m}(\mathbf{r})$ is a fermion field operator for atoms in
internal state $|\alpha m\rangle$, where $\alpha = g$ ($^1S_0$) or
$e$ ($^3P_0$) denotes the electronic state and $m=-I,\dots, I$
denotes one of the $N = 2 I + 1$ nuclear Zeeman states. The
density operators are defined as $\rho_{\alpha m}(\mathbf{r}) =
\Psi^\dagger_{\alpha m}(\mathbf{r}) \Psi_{\alpha m}(\mathbf{r})$
and $\rho_{\alpha}(\mathbf{r}) = \sum_m \rho_{\alpha
m}(\mathbf{r})$. The term $V_{\alpha}(\mathbf{r})$ describes the
external trapping potential, which we will assume to be an optical
lattice  independent of the nuclear spin: even for a relatively
deep lattice with a 100 kHz trap frequency, tensor and vector
light shifts should be well below 1 Hz \cite{Boyd2007}. $\hbar \omega_0$ is the transition energy between
$|g\rangle$ and $|e\rangle$.
Extra lasers can be used to drive transitions between $|g\rangle$ and $|e\rangle$ levels \cite{Boyd2007, Campbell2009}.  
Since we will only need these extra lasers for system preparation, we have not included the corresponding terms in the Hamiltonian.

\begin{figure}
  \begin{center}
   \includegraphics[width=80mm]{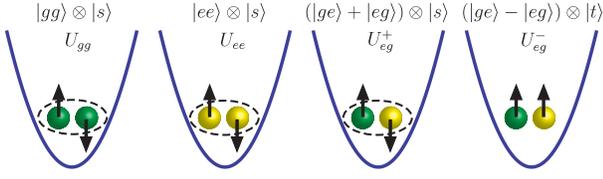}
  \end{center}
  \caption{\textbf{Interaction parameters between $g$ (green) and $e$ (yellow) atoms loaded in the lowest vibrational state of the corresponding optical lattice.} Here we assumed $I=1/2$, and the arrows indicate the $m_I=\pm 1/2$ spin states.  $|s,t\rangle$ denote the singlet and triplet nuclear spin states of the two atoms (only one of three triplet states - $|\uparrow \uparrow\rangle$ - is shown). The dashed circle represents anti-symmetrization of the nuclear spin state (i.e.~$|s\rangle$). The interaction energy $U_X$ ($X = gg, ee, eg^+, eg^-$) is proportional to the corresponding scattering length $a_X$.}
\label{inter}
\end{figure}

The interaction is characterized by four $s$-wave scattering lengths $a_X$, $X = gg, ee, eg^+, eg^-$, which define four interaction parameters $g_X = 4 \pi \hbar^2 a_X/M$, where $M$ is
atomic mass. $a_{gg}, a_{ee}$, and $a^\pm_{eg}$ are the scattering length for two atoms in the electronic state $|gg\rangle$, $|ee\rangle$, and $|\pm\rangle =(|ge\rangle
+|eg\rangle)/\sqrt{2}$, respectively. As shown in Fig.~\ref{inter}, the fermionic antisymmetry then forces the nuclear state to be symmetric for the only antisymmetric electronic state $|-\rangle$ and antisymmetric otherwise. Very few $a_X$ are known at the moment (see Supplementary Information).

The independence of each of the four scattering lengths from
the nuclear spin state is essential to the fulfillment of the SU(N)
symmetry of our model (see next Section).  This independence is a
consequence of the decoupling between nuclear and electronic
degrees of freedom exhibited  during the course of a collision
involving any combination of $g$ or $e$ states, which both have $J
= 0$. While for the $|e\rangle \equiv {}^3P_0$ atom, the
decoupling is slightly broken by the admixture with higher-lying P
states with $J \neq 0$, this admixture is very small
\cite{Boyd2007} and the resulting nuclear-spin-dependent variation of the
scattering lengths is also expected to be very small, on the order of $10^{-3}$ (see Supplementary Information). 
For $a_{gg}$, which does not involve state $|e\rangle$,
this variation 
should be even smaller ($\sim 10^{-9}$).

The interaction terms in Eq.\ (\ref{ham0}) describe the most general s-wave two-body interaction consistent with elastic collisions as far as the electronic state is concerned and with the independence of the scattering length from the nuclear spin. 
While the assumption of elasticity for $g$-$g$ and $e$-$g$ collisions is well justified, since no inelastic exit channels exist, $e$-$e$ collisions are likely to be accompanied by large losses, which means that the magnitudes of the imaginary and real parts of the $e$-$e$ scattering length are likely to be comparable  (see Supplementary Information). Therefore, we focus below on those situations where two $e$ atoms never occupy the same site.

We assume that only the lowest band in both $e$ and $g$ lattices is occupied and expand the field operators in terms of the corresponding (real) Wannier basis functions 
$\Psi_{\alpha m}(\mathbf{r}) = \sum_{j} w_{\alpha}(\mathbf{r} - \mathbf{r}_j) c_{j \alpha m}$, where $c^\dagger_{j \alpha m}$ creates an atom in internal state $|\alpha m\rangle$ at site $j$ (centered at position $\mathbf{r}_j$). Eq.\ (\ref{ham0}) reduces then to a two-orbital single-band Hubbard Hamiltonian
\begin{eqnarray}
  {H} &=&- \!\!\!\! \sum_{\langle j,i\rangle \alpha,m} \!\!\!\! J_\alpha
(c_{i\alpha m}^\dagger   c_{j\alpha m} +\textrm{h.c.}) +
  \sum_{j, \alpha} \frac{U_{\alpha \alpha}}{2} n_{j \alpha} (n_{j \alpha} - 1)\notag\\&& + V\sum_{j}
   n_{j e}  n_{jg} +V_{ex} \sum_{j,m,m' }
  {c}_{ jg m }^{\dagger}     {c}_{ jem'}^{\dagger}
  {c}^{}_{jgm'}  {c}^{}_{jem}. \label{ham}
\end{eqnarray}
%
%
Here  $J_{\alpha} = - \int d^3 \mathbf{r} w_\alpha (\mathbf{r}) (-
\frac{\hbar^2}{2 M} \nabla^2 + V_{\alpha}(\mathbf{r}))
w_\alpha(\mathbf{r}-\mathbf{r_0})$ are the tunneling energies,
$\mathbf{r_0}$ connects two nearest neighbors, h.c.~stands for Hermitian conjugate,  $n_{j \alpha m}=  c^\dagger_{j \alpha m}  c_{j \alpha m}$, and $ n_{j \alpha}=\sum_m   n_{j \alpha m}$.  The tunneling is
isotropic, which is a crucial difference between this model and
its analogues in solid state systems with orbital degeneracy \cite{Kugel1973b}.
The sum $\langle j,i\rangle$ is over pairs of nearest neighbor sites $i$, $j$.
 $V = (U^+_{eg} + U^-_{eg})/2 $
and $V_{ex} = (U^+_{eg} - U^-_{eg})/2$ describe 
the direct and exchange
interaction terms. 
The onsite interaction energies are $U_{\alpha \alpha} = g_{\alpha \alpha} \int
d^3\mathbf{r}w^4_\alpha(\mathbf{r})$ and $U^\pm_{eg} = g^\pm_{eg} \int
d^3\mathbf{r}w^2_e(\mathbf{r}) w^2_g(\mathbf{r})$. Constant terms, proportional to $\sum_j n_{j \alpha}$, 
are omitted in Eq.\ (\ref{ham}). Experimental control over the parameters in Eq.~(\ref{ham}) will allow us to manipulate the atoms (see Methods).





\section*{Symmetries of the Hamiltonian}

To understand the properties of the Hamiltonian in Eq.\
(\ref{ham}), we consider its symmetries. We define SU(2) pseudo-spin
algebra via
\begin{equation}
T^{\mu} =  \sum_j T^\mu_{j} =\frac{1}{2}  \sum_{j m \alpha \beta}  c^\dagger_{j \alpha m} \sigma^\mu_{\alpha \beta} c_{j \beta m},
\end{equation}
where  $\sigma^\mu$ ($\mu = x, y, z$) are Pauli matrices in the $\{e,g\}$ basis.
We further define nuclear-spin
permutation operators
\begin{eqnarray}
S^m_n = \sum_{j} S^m_n(j) =  \sum_{j, \alpha} S^m_n(j, \alpha) = \sum_{j, \alpha} c^\dagger_{j \alpha n} c_{j \alpha m},
\end{eqnarray}
which satisfy the SU(N) algebra $[S^m_n ,S^p_q] = \delta_{mq} S^p_n - \delta_{pn} S^m_q$, and thus generate SU(N) rotations of nuclear spins ($N = 2 I + 1$).

In addition to the obvious conservation of the total number of
atoms $n = \sum_j (n_{je} + n_{jg})$, $H$ exhibits $U(1) \times
SU(N)$ symmetry (see Methods for the discussion of enhanced
symmetries), where $U(1)$ is associated with the elasticity of
collisions as far as the electronic state is concerned ($[T^z, H]
= 0$) and SU(N) is associated with the independence of scattering
and of the trapping potential 
from the nuclear spin ($[S^m_n, H] = 0$ for all $n$,
$m$). The two-orbital SU(N)-symmetric Hubbard Hamiltonian in Eq.\
(\ref{ham}) is a generalization to $N > 2$ of its
SU(2)-symmetric counterpart \cite{Kugel1973b}
 and  to two orbitals of its single-orbital counterpart \cite{Marston1989}. 
  The SU(N) symmetry and the largely independent spin and orbital degrees of freedom are two unique features  present in alkaline-earths but absent in alkalis due to strong hyperfine interactions. 


One important consequence of SU(N) symmetry is the conservation, for any
$m$, of $S_m^m$, the total number of atoms with nuclear spin $m$. This means
that an atom with large $I$, e.g. ${}^{87}$Sr ($I = 9/2$), can reproduce the
dynamics of atoms with lower $I$ if one takes an initial state with $S_m^m = 0$ for some $m$ values.
To verify SU(N) symmetry of the
interaction experimentally, one could, thus, put two atoms in one
well in spins $m$ and $m'$ and confirm that collisions do not
populate other spin levels. This feature of SU(N) symmetry is in stark contrast to the case of weaker SU(2) symmetry, where the dependence of scattering lengths on the total spin of the two colliding particles allows
 for scattering into spin states other than $m$ and $m'$. 
We note that  although collisions are
governed by electronic interactions and obey the nuclear-spin
SU(N) symmetry, the nuclear spins  still indirectly control the
collisions via fermionic statistics and give rise to effective
spin-orbital  and spin-spin interactions.

One can alternatively implement the two-orbital Hubbard model with
two ground-state species of alkaline-earth atoms (e.g.\ ${}^{171}$Yb and ${}^{173}$Yb, or ${}^{173}$Yb and ${}^{87}$Sr). If we still
refer to them as $|g\rangle$ and $|e\rangle$, the
nuclear distinguishability and the fact that both atoms are in the ground state will result in $a_{eg}^+ = a_{eg}^-$, 
corresponding to an enhanced symmetry 
(see Methods). While experimentally more
challenging, the use of two different ground state species will
solve the problem of losses associated with collisions of
two excited state atoms and will reduce 
the (already very weak) nuclear-spin-dependence of $a_{ee}$ and $a_{eg}$.


\section*{Spin Hamiltonians}

\begin{figure}[t]
  \begin{center}
  \includegraphics[scale=0.7]{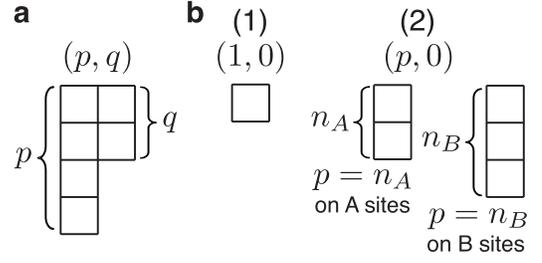}
  \end{center}
  \caption{\textbf{Young diagrams describing the irreducible representations of SU(N) on individual sites.} \textbf{a,} A general diagram consists of $n_j$ boxes arranged into at most two columns (to satisfy fermionic antisymmetry with only two orbital states) whose heights we will denote by $p$ and $q$, such that $N \geq p \geq q$ and $p+q = n_j$. See Supplementary Information 
for a brief review of Young diagrams. \textbf{b,} The Young diagrams for the two special cases discussed in the main text: \textbf{(1)} $(p,q) = (1,0)$ and \textbf{(2)} $(p,q) = (p,0)$ on a bipartite lattice.}
\label{Youngfigure}
\end{figure}

One  of the simplest interesting limits of Eq.\ (\ref{ham}) is the  strongly interacting regime ($J/U\ll 1$)  where the Hilbert space is restricted to a given energy manifold of the $J_g = J_e = 0$ Hamiltonian (with a fixed number of atoms on each site),
and tunneling is allowed only virtually, giving rise to an effective spin (and pseudo-spin) Hamiltonian. 
Single-site energy manifolds can  be classified according to the number of atoms $n_j = n_{jg} + n_{je}$, the pseudo-spin component $T_j^z$, and the spin symmetry (SU(N) representation) described  by a Young diagram. 
As shown in Fig.\ \ref{Youngfigure}a, each diagram  consists of $n_j$ boxes and at most two columns 
of heights $p$ and $q$, representing two sets of antisymmetrized indices.   

The $U(1) \times SU(N)$ symmetry of Eq.\ (\ref{ham}) restricts the order $J^2$ spin Hamiltonian to the form
\begin{eqnarray} \label{spin}
&& H_{(p,q)} = \sum_{\langle i,j \rangle, \alpha } \Big[ \kappa_\alpha^{ij} n_{i \alpha} n_{j \alpha} + \lambda_\alpha^{ij} S_m^n (i,\alpha) S_n^m(j, \alpha) \Big]  \nonumber  \\
&&+ \sum_{\langle i,j \rangle} \Big[\kappa^{ij}_{ge} n_{ig} n_{je} + \lambda_{ge}^{ij} S_m^n (i,g) S_n^m(j,e) \nonumber \\
&&+ \tilde \kappa^{ij}_{ge} S_{gm}^{em}(i) S_{en}^{gn}(j) + \tilde \lambda_{ge}^{ij} S_{gm}^{en}(i) S_{en}^{gm}(j) + \{i\leftrightarrow j\}\Big],
\end{eqnarray}
%
%
where the sum over $n$ and $m$ is implied in all but the $\kappa$ 
terms and $S^{\alpha m}_{\beta n}(j) = c^\dagger_{j \beta n} c_{j \alpha m}$. $\{i\leftrightarrow j\}$ means that all 4 preceding terms are repeated with $i$ and $j$ exchanged. 
The coefficients $\kappa$, $\lambda$, $\tilde \kappa$, and $\tilde \lambda$ are of order $J^2/U$ with the exact form determined by what single-site energy manifolds we are considering. $\kappa$ terms describe nearest neighbor repulsion or attraction, while $\lambda$, $\tilde \kappa$, and $\tilde \lambda$ terms describe nearest neighbor exchange of spins, pseudo-spins, and complete atomic states, respectively. Without loss of generality, $\kappa^{ij}_\alpha = \kappa^{ji}_\alpha$ and $\lambda^{ij}_\alpha = \lambda^{ji}_\alpha$. 
In many cases (e.g.\ case (2) 
below), the Hilbert space, which $H_{(p,q)}$ acts on, has $n_{ie}$ and $n_{ig}$ constant for all $i$, which not only forces $\tilde \kappa^{ij}_{ge} = \tilde \lambda_{ge}^{ij} = 0$ but also allows one to ignore the constant $\kappa_\alpha^{ij}$ and $\kappa_{ge}^{ij}$ terms. We now discuss two special cases of $H_{(p,q)}$ shown in Fig.\ \ref{Youngfigure}b. A third case, $(p,q) = (1,1)$, which reduces for $N = 2$ to the spin-1 Heisenberg antiferromagnet is discussed in the Supplementary Information.

\textbf{(1)} In the case of one atom per site, 
$(p,q) = (1,0)$.  $H_{(p,q)}$ is then
a generalization to arbitrary N of the SU($N=2$) Kugel-Khomskii model \cite{Kugel1973b, Tokura2000}, and we rewrite it  as (see Supplementary Information)
\begin{eqnarray}
&&  {H}_{(1,0)} =  \sum_{\langle i,j\rangle} \Big[ 2 (\tilde \kappa_{ge} + \tilde \lambda_{ge} S^2_{ij}) ( {T}_i^x {T}_j^x+ {T}_i^y {T}_j^y) +  \lambda_{ge} S^2_{ij} \notag \\
&& - [A+ B S^2_{ij}]( {T}_i^z {T}_j^z+\frac{1}{4}) + h (1-S^2_{ij}) ( {T}_i^z + {T}_j^z) \Big]   \label{H10},
\end{eqnarray}
%
where $S^2_{ij} = \sum_{m n} S_m^n(i) S_n^m(j)$  is $+1$ ($-1$) for a symmetric (antisymmetric) spin state,
$A=2 \kappa_{ge} - \kappa_e - \kappa_g$, $B= 2 \lambda_{ge} + \kappa_e + \kappa_g$, and $h= (\kappa_e - \kappa_g)/2$.
The $N = 2$ Kugel-Khomskii Hamiltonian is used to model the spin-orbital interactions (not to be confused with relativistic spin-orbit coupling) 
in  transition metal oxides with perovskite structure  \cite{Tokura2000}. Our implementation allows to realize clean spin-orbital interactions unaltered by lattice and Jahn-Teller distortions present in solids \cite{Tokura2000}.



\begin{figure}[t]
  \begin{center}
   \includegraphics[scale = 1]{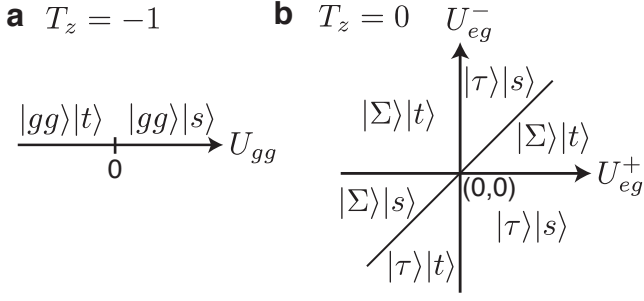}
  \end{center}
  \caption{\textbf{The ground-state phase diagram for the SU(N=2) Kugel-Khomskii model restricted to two wells, left (L) and right (R).} \textbf{a,} The phase diagram for $T_z = -1$ (two $g$ atoms). $|gg\rangle = |gg\rangle_{LR}$. 
  $|s\rangle$ and $|t\rangle$ are spin singlet and triplet states, respectively. \textbf{b,} The phase diagram for $T_z = 0$ (one $g$ atom and one $e$ atom). $|\Sigma\rangle=\frac{1}{\sqrt{2}}(|e g\rangle_{LR}-|g e\rangle_{L R})$ and  $|\tau\rangle=\frac{1}{\sqrt{2}}(|e g\rangle_{L R}+|g e\rangle_{L R})$ are  anti-symmetric and symmetric orbital states, respectively. See Supplementary Information for a detailed discussion of both of these diagrams.}
\label{KKfigure}
\end{figure}

To get a sense of the competing spin and orbital orders \cite{Arovas1995,Li1998,Pati1998} characterizing $H_{(1,0)}$, we consider the simplest case of only two sites ($L$ and $R$) and $N = 2$ (with spin states denoted by $\uparrow$ and $\downarrow$). To avoid losses in $e$-$e$ collisions, we set $U_{ee} = \infty$ (see Supplementary Information). 
The double-well ground-state phase diagram for $T^z = 1$ (two $e$ atoms) is then trivial, while the $T^z = -1$ (two $g$ atoms) and $T^z = 0$ (one $g$ atom and one $e$ atom) diagrams are shown in Fig.~\ref{KKfigure}.
One can see that, depending on the signs and relative magnitudes of the interactions, various combinations of ferromagnetic (triplet) and antiferromagnetic (singlet) spin and orbital orders are favored. In the Methods, we propose a double-well experiment  along the lines of Ref.\ \cite{Trotzky2008} to probe the spin-orbital interactions giving rise to the $T^z = 0$ diagram in Fig.~\ref{KKfigure}b. Multi-well extensions of this experiment may shed light on the model's many-body phase diagram, which has 
been studied 
for $N = 2$ and mostly at mean-field level or in special cases, such as in one dimension or in the presence of enhanced symmetries (see e.g.~\cite{Arovas1995, Pati1998,Li1998}).

\textbf{(2)} In order to study SU(N) spin physics alone, we consider the
case of $g$ atoms only. 
On a
bipartite lattice with sublattices A and B, 
we choose A sites to have $n_A < N$ atoms [$(p,q) = (n_A,0)$] and B sites to
have $n_B < N$ atoms [$(p,q) = (n_B,0)$].  This setup can be
engineered in cold atoms by using a superlattice to adjust the
depths of the two sublattices 
favoring a higher filling factor in
deeper wells.
 $H_{(p,q)}$ then reduces to
  \begin{eqnarray} \label{Hp0}
H_{(p,0)} = \frac{2 J_g^2 U_{gg}}{U_{gg}^2-(U_{gg}(n_A-n_B)+\Delta)^2} \sum_{\langle i,j \rangle}  S^2_{ij}, 
\end{eqnarray}
where $\Delta$ is the energy offset between adjacent lattice sites. The coupling constant can be made either positive (antiferromagnetic) or negative (ferromagnetic) depending on the choice of parameters \cite{Trotzky2008}.
 Three body recombination processes 
 will likely limit the lifetime of the atoms when $n_j \geq 3$ (see Supplementary Information). 

\begin{figure}[t]
  \begin{center}
   \includegraphics[scale = 1]{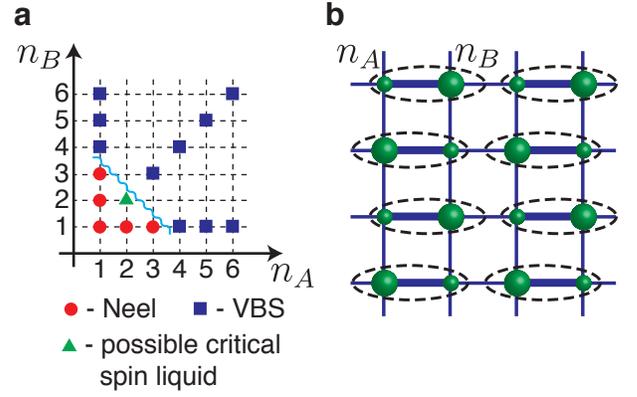}
  \end{center}
  \caption{\textbf{Probing the phases of the SU(N) antiferromagnet on a 2D square lattice.} \textbf{a} shows the phase diagram for the case $n_A + n_B = N$. Some points on this diagram have been explored in earlier numerical studies \cite{Harada2003,Assaad2005,Paramekanti2007} 
  and are marked according to the ground state obtained: Neel (circles), columnar-valence-bond solid (VBS) [shown schematically in \textbf{b}] (squares), and possibly critical spin liquid (triangle)  \cite{Assaad2005,Paramekanti2007}. 
  Since for sufficiently large $N$ quantum fluctuations tend to destabilize long-range  magnetic ordering,  it is likely that VBS ordering  characterizes the ground state for all $N > 4$ (i.e.\ above the wavy line).}
\label{SUNfigure}
\end{figure}

We focus on the 2D square lattice in the antiferromagnetic regime. The case $n_A + n_B = N$ shares with the SU(2) Heisenberg model the crucial
property that two adjacent spins can form an SU(N) singlet, and has thus been studied extensively as a large-N generalization of
SU(2) magnetism \cite{Marston1989, Read1989b}.
Fig.\ \ref{SUNfigure}a shows the expected phase
diagram for the case $n_A + n_B = N$, which features Neel (circles), valence-bond-solid (VBS) (squares) [Fig.\ \ref{SUNfigure}b], and possible critical spin liquid (triangle)  \cite{Assaad2005,Paramekanti2007} ground states. 
To access various ground states of the system, the initial state must be carefully prepared so that the conserved quantities $S^m_m$ take values appropriate for these ground states. Another interesting and experimentally relevant case, $n_A = n_B \neq N/2$, which can also exhibit spin liquid and VBS-type ground states, is discussed in the Supplementary Information and in Ref.~\cite{Hermele2009}.


Since one can vary $N$ just by choosing the number of initially populated Zeeman levels (\emph{e.g.} via a combination of optical pumping and coherent manipulation), alkaline-earth atoms offer a unique arena to
probe the phase diagram of $H_{(p,0)}$, including exotic phases such as VBS [Fig.~\ref{SUNfigure}b], 
as well as
competing magnetically ordered states. We propose to load a band insulator of $N$ $g$ atoms per site, then slowly split each well into two to form an array of independent SU(N) singlets in a pattern shown in Fig.~\ref{SUNfigure}b. 
The intersinglet tunneling rate should then be adiabatically increased up to the intrasinglet tunneling rate.
As $N$ increases, the magnetic or 
singlet nature of the state can be probed by measuring the Neel order parameter (see the description of the Kugel-Khomskii double-well experiment in the Methods) 
 and  spin-spin correlations via noise spectroscopy in the time of flight \cite{Altman2004} (which directly measures $\sum_{i,j}\langle S_n^m(i,g) S_m^n(j,g)\rangle e^{I Q (i-j)}$). 


\section*{The Kondo lattice model (KLM)}

The SU(N) Kondo
lattice model (KLM) \cite{Coqblin1969,Coleman1983} 
is another example of the rich  physics, beyond the Mott regime, which could be simulated with alkaline-earth atoms. 
  The KLM  is  one of  the canonical models
used to study strongly correlated electron systems, such as
manganese oxide perovskites \cite{Tokura2000book} and rare
earth and actinide compounds classed as heavy fermion materials \cite{Coleman2007}. 

\begin{figure}[t]
  \begin{center}
   \includegraphics[scale = 0.8]{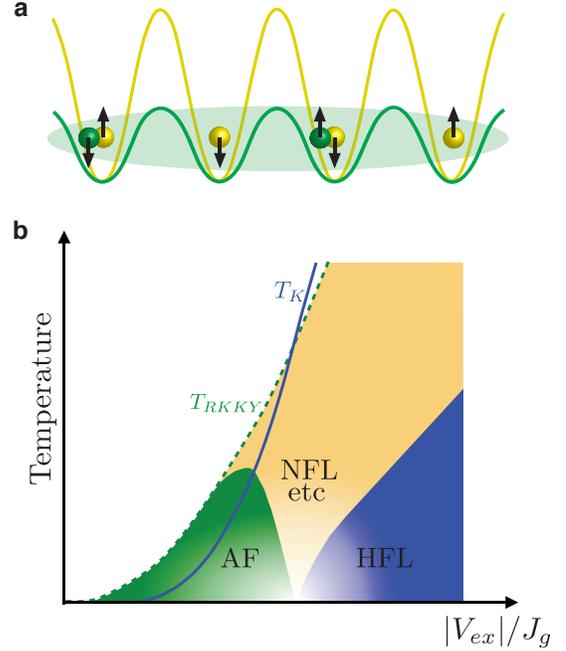}
     \end{center}
\caption{\textbf{Kondo lattice model for the case $N = 2$.} \textbf{a,} The schematic of the setup. $g$ atoms are green; $e$ atoms are yellow; the spin basis is $\left\{\uparrow, \downarrow\right\}$. \textbf{b,}
Schematic  representation of the competition between RKKY magnetism vs Kondo singlet formation in the SU(2) AF KLM 
(see \cite{Doniach1977,Coleman2007,Gegenwart2008} and references therein). In this model, the localized spin-$1/2$ $e$ atoms couple  antiferromagnetically to the delocalized $g$ atoms, via an on-site exchange interaction proportional to $V_{ex}$. This coupling  favors the formation of localized Kondo singlets between $e$ and $g$ atoms, with characteristic energy scale $k_B T_K \sim J_{g} \exp(-c J_g/|V_{ex}|)$, with $c$  a dimensionless constant of order one  \cite{Coleman2007}. 
On the other hand, the $g$ atoms can mediate long-range RKKY  interactions between the $e$ atoms,  giving rise to magnetic order (which can be antiferromagnetic (AF) or ferromagnetic depending on the density of $g$ atoms), where the characteristic energy is $k_B T_{RKKY} \sim V_{ex}^2/J_g$.  
 The competition between Kondo effect and RKKY magnetism leads to very rich physics.
 For small values of $|V_{ex}|/J_g$, the RKKY interaction is dominant and the system orders magnetically. At intermediate values of $|V_{ex}|/J_g$, the  energy scales $T_K$ and $T_{RKKY}$ are of comparable strength, and a variety of novel quantum phenomena are expected to arise, including quantum criticality and non-Fermi liquid (NFL) physics \cite{Coleman2007,Gegenwart2008}.  With further increase of the $|V_{ex}|/J_g$ coupling, magnetic order is  suppressed, the localized $e$ atoms  become screened into singlet states and  melt  into the $g$-atom Fermi sea, forming  the so called heavy Fermi liquid state (HFL). The large Fermi volume \cite{Oshikawa2000}, which is set by the total number of $g$ atoms plus $e$ atoms, can be directly probed by measuring the
momentum distribution  via time of flight imaging. }
 \label{Kondo}
\end{figure}

For its implementation with cold atoms (for $N = 2$, see also Refs.~\cite{Duan2004,Paredes2005}), we propose to put one $e$
atom (localized spin) per site in a deep lattice such that $J_e
\ll U_{ee}$, so that we can set $J_e = 0$ and $  n_{je} = 1$ for
all $j$ in Eq.\ (\ref{ham}). We also suppose that we can set $U_{gg} = 0$, e.g.~by taking a very shallow $g$-lattice (see Fig.\ \ref{Kondo}a).
The resulting Hamiltonian is the SU(N) KLM
\cite{Coqblin1969,Coleman1983}
\begin{eqnarray}
  {H}_{KLM}  \!\!\! &=&  \!\!\! -  \!\!\!\! \sum_{\langle j,i\rangle m}  \!\!\!\! J_g   (c_{i g m}^\dagger   c_{jg m}\!+\!\textrm{h.c.})  \!+\! V_{ex} \!\!\!\!\! \sum_{j,m,m' }  \!\!\!\!  {c}_{ jgm }^{\dagger}     {c}_{ jem'}^{\dagger}
   {c}^{}_{jgm'}  {c}^{}_{jem}. \label{KLM}
 \end{eqnarray}
The magnitude of $V_{ex}$ can be adjusted by shifting the $e$ and $g$ lattices relative to each other \cite{Daley2008}.

The properties of the SU(N) KLM depend crucially on the sign of the
exchange interaction. For concreteness, we focus on the antiferromagnetic (AF) case ($V_{ex} < 0$), which favors formation of spin-antisymmetric states (singlets, for $N=2$)
between mobile fermions and localized spins.  This regime describes the physics of heavy
fermion materials \cite{Coleman2007}, 
and, in the case
of a single localized spin, gives rise to the Kondo effect.

In the limit $|V_{ex}| \ll J_g$, $g$ atoms mediate long-range RKKY interactions \cite{Ruderman1954}
between localized spins and tend to induce magnetic ordering (antiferromagnetic or ferromagnetic depending on the density of $g$ atoms) of the latter, at least for $N = 2$.
The engineering of RKKY interactions  can be tested in an array of isolated double wells (see Methods).
At intermediate and large $|V_{ex}|$, the formation of Kondo singlets dominates the RKKY interaction and favors a magnetically disordered heavy Fermi liquid (HFL) ground state with a significantly enhanced effective quasiparticle mass  (see Fig.~\ref{Kondo}b). The competition between  RKKY interactions and the Kondo effect in the regime where both are comparable is subtle, and the resulting phases and phase transitions \cite{Coleman2007,Gegenwart2008} are not well-understood. 
Ultracold alkaline-earth atoms provide a promising platform to study these phases and phase transitions.

 In the large-$N$ limit \cite{Coqblin1969,Coleman1983}, the SU(N) HFL can be controllably studied, and $1/N$ expansions have   successfully reproduced the experimentally observed properties of the HFL.  However,
very little is known about the SU(N) model outside the HFL regime. Several very interesting parameter regimes in this domain can be directly probed with our system, as discussed in the Methods.

\section*{Experimental Accessibility}

The phenomena described in this manuscript can be probed with experimental systems under development.
Indeed, we show in the Methods that SU(N)-breaking terms are sufficiently weak, and here we discuss the temperature requirements.

The key energy scale in the spin Hamiltonians [Eq.~(\ref{spin})] is the superexchange energy $J^2/U$, while the RKKY energy scale is $k_B T_{RKKY} \sim V_{ex}^2/J_g$.  In their region of validity ($J < U$ and $|V_{ex}| < J_g$, respectively), these energy scales are limited from above by the interaction energy ($U$ and $|V_{ex}|$, respectively), which typically corresponds to temperatures $T \lesssim 100$nK \cite{Trotzky2008}. Thanks to the additional cooling associated with certain adiabatic changes \cite{Hofstetter2002, Werner2005}, $T \sim 10$nK and the Mott insulating regime have already been achieved with fermionic alkali atoms \cite{Schneider2008}, and are therefore expected to be achievable with fermionic alkaline-earths, as well (a bosonic alkaline-earth Mott insulator has already been achieved \cite{Fukuhara2009}). Furthermore, the requirement to reach $k_B T$ smaller than $J^2/U$ or $V_{ex}^2/J_g$ can often be relaxed. First, the double-well experiments, such as the ones discussed in the Methods in the contexts of the Kugel-Khomskii and the Kondo lattice models, are performed out of thermal equilibrium, and can, thus, access energy scales far below the temperature of the original cloud \cite{Trotzky2008}. Second, for SU(N) antiferromagnets, the energy range between $J^2/U$ and $N J^2/U$ may also exhibit intriguing physics: in this regime, SU(N) singlets, which require $N J^2/U$ energy to break, stay intact but can diffuse around. 
Finally, in the $V_{ex} < 0$ Kondo lattice model, exotic heavy Fermi liquid behavior is expected when $J_g \lesssim |V_{ex}|$ and the temperature is below the Kondo temperature, i.e.~$k_B T \lesssim J_{g} \exp(-c J_g/|V_{ex}|)$ with $c$ is a dimensionless constant of order one  \cite{Coleman2007}. 
Thus, with $J_g$ chosen to be on the order of $|V_{ex}|$, $k_B T$ as high as $\sim |V_{ex}|$ may be sufficient.

\section*{Outlook}

The proposed experiments 
should be regarded  as  bridges aiming to
connect  well-understood  physics to the complex and poorly
understood behavior of strongly correlated  systems. It is
important to emphasize that, except for the one dimensional case,
the phase diagram of most of the models considered is  only known at mean
field level or numerically in reduced system sizes. Therefore,
their experimental realization in  clean and controllable
ultracold atomic systems  can provide major advances. 

Our proposal motivates other new lines of research. Ultracold bosonic or fermionic diatomic molecules \cite{Ni2008} may give rise to similar SU(N) models with large $N$ and with the possibility of long-range interactions. Ions with alkaline-earth-like structure, such as Al$^+$ could also be considered in this context. It would also be interesting to explore the possibility of realizing topological phases with SU(N) models for applications in topological quantum computation \cite{Hermele2009}. 
Beyond quantum magnetism, 
the fact that the formation of SU(N) singlets requires $N$ partners
might  give rise to novel exotic 
types of superfluidity and novel types of
BCS-BEC crossover \cite{Rapp2008}.
Practical applications of our 
Hubbard model, such as the calculation of the collisional frequency shift in atomic clocks \cite{Rey2009}, can also be foreseen.

\vspace{0.4in}

\textit{Note added in proof.} After the submission of this paper, a theoretical study of the SU(6)-symmetric ${}^{173}$Yb system was reported \cite{Cazalilla2009}.

\section*{Methods}

\subsection*{Experimental tools available for alkaline-earth atoms}

Many experimental tools, such as tuning the interaction strength by adjusting laser intensities \cite{Trotzky2008}, are common to both alkali and alkaline-earth atoms.
There are, however, some experimental tools specific to alkaline earths; we review them in this Section.

First, a combination of optical pumping \cite{Campbell2009}
and direct coherent manipulation of the $|g\rangle-|e\rangle$ transition in the presence of a magnetic field \cite{Boyd2007, Campbell2009} 
can be used \cite{Gorshkov2009} to prepare any desired single-atom state within the 2 (2 I + 1)-dimensional manifold with basis $|\alpha m\rangle$, where $\alpha = g$ or $e$ and $m = -I, \dots, I$. This coherent manipulation can also be used to exchange quantum information between nuclear spin states and electronic states. Second, by using far-detuned probe light or a large magnetic field to decouple the electronic angular momentum $J$ and the nuclear spin $I$, the electronic $|g\rangle-|e\rangle$ degree of freedom can be measured by collecting fluorescence without destroying the nuclear spin state \cite{Gorshkov2009}. Fluorescence measurement of the nuclear spins can be achieved by mapping nuclear spin states onto electronic states \cite{Gorshkov2009,Daley2008}: for example, for a spin-$1/2$ nucleus, a $\pi$ pulse between $|g, m = 1/2\rangle$ and $|e, m = -1/2\rangle$ allows one to accomplish a swap gate between the nuclear $\{1/2,-1/2\}$ qubit and the electronic $\{e,g\}$ qubit. Single-site spatial resolution during the coherent manipulation and fluorescence measurement can be achieved using magnetic field gradients \cite{Daley2008} or dark-state-based techniques 
\cite{Gorshkov2008,Gorshkov2009}
that rely on an auxiliary laser field whose intensity vanishes at certain locations. Third, an appropriate choice of laser frequencies allows one to obtain independent lattices for states $g$ and $e$ \cite{Daley2008}. Finally, optical Feshbach resonances \cite{Ciurylo2005} 
may be used to control scattering lengths site-specifically and nearly instantaneously.

\subsection*{Enhanced Symmetries}

While in the general case, our Hubbard model [Eq.\ (\ref{ham})]
satisfies $U(1) \times SU(N)$ symmetry, for particular choices of
parameters, higher symmetry is possible. In particular, if $J_g =
J_e$ and the interaction energies for all states within the pseudo-spin
triplet are equal  ($U_{gg} = U_{ee} = U_{eg}^+$), the full SU(2)
symmetry (not just U(1)) in the pseudo-spin space is satisfied.
Alternatively, if $V_{ex} = 0$, then both $S_n^m(i,g)$ and
$S_n^m(i,e)$ generate SU(N) symmetries resulting in the overall
$U(1) \times SU(N) \times SU(N)$ symmetry. Finally, if both
conditions are satisfied, i.e.\ all four $U_X$ are equal and $J_g = J_e$, then $H$
satisfies the full SU(2N) symmetry ($2N$ can be as high as 20)
generated by
\begin{equation}
S^{\alpha m}_{\beta n} = \sum_j S^{\alpha m}_{\beta n}(j) = \sum_j c^\dagger_{j \beta n} c_{j \alpha m},
\end{equation}
in which case the interaction reduces to $\frac{U}{2} \sum_{j} n_j
(n_j -1)$, where $n_j = n_{jg} + n_{je}$.

In the case when $|e\rangle$ and $|g\rangle$ correspond to two ground states of two different atoms (with nuclear spin $I_e$ and $I_g$, respectively), we will have $a_{eg}^+ = a_{eg}^-$ (i.e\ $V_{ex} = 0$), which is equivalent to imposing $U(1) \times SU(N_g = 2 I_g + 1) \times SU(N_e = 2 I_e+ 1)$  symmetry, where $SU(2 I_\alpha + 1)$ is generated by $S_n^m(i,\alpha)$. While for $I_g \neq I_e$, the $m$ index in $c_{j \alpha m}$ will run over a different set of values depending on $\alpha$, the Hubbard Hamiltonian will still have the form of Eq.\ (\ref{ham}) (except with $V_{ex} = 0$). If one further assumes that $J_g = J_e$ and $U_{gg} = U_{ee} = U_{eg}$, the interaction satisfies the full $SU(N_g + N_e)$ symmetry. It is worth noting that for the case of two different ground state atoms, this higher symmetry is easier to achieve than for the case of two internal states of the same atom, since $a_{eg}^+ = a_{eg}^-$ automatically. Thus, in particular, it might be possible to obtain $SU(18)$ with ${}^{87}$Sr ($I = 9/2$) and ${}^{43}$Ca ($I = 7/2$) simply by adjusting the intensities of the two lattices (to set $J_g = J_e$ and $U_{gg} = U_{ee}$) and then shifting the two lattices relative to each other (to set $U_{eg} = U_{gg}$).

Enhanced symmetries of the Hubbard model [Eq.\ (\ref{ham})] are
inherited by the spin Hamiltonian [Eq.\ (\ref{spin})]. In
particular, imposing $SU(2) \times SU(N)$ instead of $U(1) \times
SU(N)$ forces $\kappa_{ge}^{ij} = \kappa_{ge}^{ji}$, $\tilde \kappa_{ge}^{ij} =
\tilde \kappa_{ge}^{ji}$, $\kappa^{ij}_g = \kappa^{ij}_e = \kappa^{ij}_{ge} + \tilde
\kappa^{ij}_{ge} \equiv \kappa^{ij}$, $\lambda_{ge}^{ij} = \lambda_{ge}^{ji}$, $\tilde
\lambda_{ge}^{ij} = \tilde \lambda_{ge}^{ji}$, $\lambda^{ij}_g = \lambda^{ij}_e =
\lambda^{ij}_{ge} + \tilde \lambda^{ij}_{ge} \equiv \lambda^{ij}$. Alternatively,
imposing $U(1) \times SU(N) \times SU(N)$ forces $\tilde
\kappa^{ij}_{ge} = \lambda^{ij}_{ge} = 0$. Finally, imposing the full SU(2N)
forces the satisfaction of both sets of conditions, yielding
\begin{eqnarray}
H &= &  \sum_{\langle i,j\rangle} \Big[ \kappa^{ij} n_{i}
n_{j} + \lambda^{ij} S_{\alpha m}^{\beta n} (i) S_{\beta n}^{\alpha
m}(j) \Big],
\end{eqnarray}
which is, of course, equivalent to restricting Eq.\ (\ref{spin})
to $g$-atoms only and extending labels $m$ and $n$ to run over
$2N$ states instead of $N$.

\subsection*{Double-well Kugel-Khomskii and RKKY experiments}

In the main text and in the following Methods Section, we discuss the open questions and previously unexplored regimes associated with the SU(N) Kugel-Khomskii and Kondo lattice models (KLM) that can be studied with ultracold alkaline-earth atoms. As a stepping stone toward these many-body experiments, we propose in this Section two proof-of-principle experiments in an array of isolated double wells with $N = 2$  (with the spin basis $\left\{\uparrow, \downarrow\right\}$): one to probe the spin-orbital interactions of the Kugel-Khomskii model and one to probe the RKKY interactions associated with KLM.

We first propose an experiment  along the lines of Ref.\ \cite{Trotzky2008} to probe the spin-orbital interactions giving rise to the $T^z = 0$ diagram in Fig.~\ref{KKfigure}b. In the Supplementary Information, we describe how to prepare an array of independent double wells in the state $|e,\uparrow\rangle_L|g,\downarrow\rangle_R$, which is a superposition of the four eigenstates featured in Fig.~\ref{KKfigure}b. The energies of these four eigenstates [Eqs.~(\ref{kksupp1}-\ref{kksupp2})] can be extracted from the Fourier analysis of the population imbalance as a function of time: $\Delta N(t)=n_{eR}+n_{gL}-n_{gR}-n_{eL} = -\cos\left[\frac{4 t J_e J_g}{\hbar U_{eg}^-}\right] - \cos\left[\frac{4 t J_e J_g}{\hbar U_{eg}^+}\right]$. $\Delta N$ can be measured by combining the dumping technique, band mapping, and Stern-Gerlach filtering of Ref.~\cite{Trotzky2008} with the use of two probe laser frequencies to distinguish between $|g\rangle$ and $|e\rangle$.

We now turn to the double-well experiment aimed at probing RKKY interactions. After preparing the state $\frac{1}{\sqrt{2}}(|g,\downarrow\rangle_L+|g,\downarrow\rangle_R)|e,\downarrow\rangle_L|e,\uparrow\rangle_R$ (see Supplementary Information for how to prepare this state), we propose to monitor the Neel order parameter for the $e$ atoms,  $N_{ez} = \frac{1}{2}[n_{e\uparrow L}-n_{e\downarrow L}-(n_{e\uparrow R}-n_{e\downarrow R})]$. In the limit $|V_{ex}| \ll J_g$, $N_{ez}(t) = -\frac{1}{3} \cos \left(\frac{V_{ex} t}{\hbar}\right) - \frac{2}{3} \cos \left(\frac{V_{ex} t}{2 \hbar} - \frac{3 V_{ex}^2 t}{8 J_g \hbar}\right)$ [in the Supplementary Information, we present the plot of $N_{ez}(t)$ for $V_{ex} = - J_g/10$]. It exhibits fast oscillations with frequency $\sim V_{ex}$, modulated by an envelope of frequency $\sim V_{ex}^2/J_g$ induced by RKKY interactions. 
In order to probe RKKY interactions only, it is important to suppress super-exchange $\sim\!\! J_{e}^2/U_{ee}$ and thus to choose $J_e/U_{ee}$ small.  To study the full spatial dependence of RKKY interactions, one must of course go beyond the double-well setup. We also note that recent experiments using alkali atoms  populating the lowest two vibrational levels of a deep optical lattice have measured  the local singlet-triplet splitting induced by $V_{ex}$ \cite{Anderlini2007}.

\subsection*{Physics accessible with the alkaline-earth Kondo lattice model}

The alkaline-earth atom realization of the AF KLM is well-suited to access a number of parameter regimes that are out of reach in solid state materials.
One example is the one dimensional (1D) limit, since, to our knowledge, real solid state materials
exhibiting KLM physics are restricted to 2D or 3D.
Another example is the   regime of large Kondo exchange ($|V_{ex}| \gg J_g$), which is interesting even for $N = 2$.
In this limit the system is well described by  the  $U\to \infty$ Hubbard model \cite{Tsunetsugu1997} by identifying the Kondo singlets  with  empty sites (holes)  and the unpaired localized spins with   hard core   electrons. From this mapping,  possible ferromagnetic ordering is expected at small hole concentration (small $n_g$), however the stability of this phase
 for increasing hole concentration and finite $|V_{ex}|$ values remains unknown.
 For general $N$, in the extreme limit $J_g = 0$, the ground state is highly degenerate:   for any distribution of the $g$ atom density $n_{j g} < N$, there is a ground state (with further spin degeneracy), where on each site the spins combine antisymmetrically to minimize the exchange interaction.  Lifting of such extensive degeneracies often leads to novel ground states; this will be addressed in future studies using degenerate perturbation theory in $J_g / V_{ex}$.  For $N > 2$, AF SU(N) spin models have a different kind of extensive degeneracy, which was argued to destroy antiferromagnetism and to lead to non-magnetic spin liquid and VBS-like ground states \cite{Hermele2009}.  Similar expectations are likely to apply to the KLM at small $|V_{ex}| / J_g$, where the $N = 2$ antiferromagnetism may give way to situations where the localized spins form a non-magnetic state that is effectively decoupled from the mobile fermions \cite{Senthil2003}.

Even though we have set $U_{g g}$ to zero  in Eq.~(\ref{KLM}), it can be tuned,  for example, by adjusting the $g$-lattice depth and can give rise to interesting physics. For example, the $n_g = 1$ case, which is known to be for  $N=2$ either an antiferromagnetic insulator or a Kondo insulator depending on the ratio $|V_{ex}|/J_g$ \cite{Assaad1999}, will become for large enough $U_{g g}$ and $N > 2$ a Mott insulator, because the two atoms on each site cannot combine to form an SU(N) singlet.  If $n_g$ is reduced from unity, the doping of this Mott insulator can be studied, and it will be
interesting to understand how this physics, usually associated with cuprate superconductors, is related to the other ground states of the KLM, usually associated with heavy fermion compounds.

\section*{Experimental Accessibility}

Immediate experimental accessibility makes our proposal particularly appealing.  Having shown in the main text that the temperature requirements of our proposal are within reach of current experimental systems, here we show that the nuclear-spin dependence of interaction energies is sufficiently weak to keep the SU(N) physics intact.

In the Supplementary Information,  nuclear-spin-dependent variation in the interaction energies is estimated to be $\Delta U_{gg}/U_{gg} \sim 10^{-9}$ and $\Delta U_{ee}/U_{ee} \sim \Delta U^\pm_{eg}/U^\pm_{eg} \sim 10^{-3}$. Since the scale of SU(N) breaking is at most $\Delta U$, a very conservative condition for 
the physics to be unaffected by SU(N) breaking is that all important energy scales are greater than $\Delta U$. In particular, in the spin models with more than one atom per site, the condition is $\Delta U \ll J^2/U$, 
which can be satisfied simultaneously with $J \ll U$ even for $\Delta U/U \sim 10^{-3}$. With one atom per site, the SU(N) breaking scale is not $\Delta U$ but rather $(J/U)^2 \Delta U$, which relaxes the condition to the immediately satisfied $\Delta U/U \ll 1$. Similarly, in the Kondo lattice model, the conditions $\Delta V_{ex} \ll J, |V_{ex}|$ can be satisfied for $\Delta V_{ex}/|V_{ex}| \sim 10^{-3}$.

\section*{Acknowledgments}

We gratefully acknowledge conversations with M.~M.~Boyd, 
A.~J.~Daley, S.~F\"olling, W.~S.~Bakr, J.~I.~Gillen, L.~Jiang, G.~K.~Campbell, and Y.~Qi. This work was supported by NSF, CUA, DARPA, AFOSR MURI, NIST.

\section*{Author contributions}


All authors contributed extensively to the work presented in this paper.

\section*{Additional information}

Supplementary information accompanies this paper on www.nature.com/naturephysics. Reprints and permissions information is available online at http://npg.nature.com/reprintsandpermissions. Correspondence and requests for materials should be addressed to A.V.G.

\newpage
\newpage

\section*{SUPPLEMENTARY ONLINE MATERIALS}

\subsection*{Known Scattering Lengths}

Very few scattering lengths $a_X$ ($X = gg, ee, eg^+, eg^-$) between $g$ ($^1S_0$) and $e$ ($^3P_0$) states of alkaline-earth-like atoms are known at the moment.
$a_{gg}$ is known for all isotopic combinations of Yb~\cite{Enomoto2008X} and Sr \cite{Escobar2008X}. Estimates of $a_{ee}$ for ${}^{88}$Sr \cite{Traverso2009X} and of 
$a^-_{eg}$ for ${}^{87}$Sr \cite{Campbell2009X} also exist. Finally, there is a proposal describing how to measure $a^+_{eg}$ via clock shifts \cite{Rey2009X}.

\subsection*{Nuclear-Spin Independence of the Scattering Lengths}

Independence of scattering lengths from the nuclear spin is a key assumption of the paper. This feature 
allows us to obtain SU(N)-symmetric models with $N$ as large as 10 and distinguishes alkaline-earth atoms from alkali atoms, which can exhibit at most an SO(5) symmetry \cite{Wu2003X,Wu2005X,Chen2005X,Wu2006X}, a symmetry that is weaker than SU(4). The assumption of nuclear-spin independence of scattering lengths is consistent with recent experiments, where - within experimental precision - the clock shift does not depend on how the Zeeman levels are populated \cite{Boyd2007bX, Ludlow2008X}. In this Section, we present the theoretical justification of this assumption.

Direct magnetic dipole-dipole coupling between the nuclear spins of two atoms sitting on the same site of an optical lattice is negligible: even for two magnetic dipole moments as large as 10 nuclear magnetons at a distance of 10 nm (which is significantly smaller than the confinement typically achieved in optical lattices \cite{Trotzky2008X}), the interaction energy still corresponds to a frequency smaller than one Hertz. Therefore, nuclei can affect the collisions only via the electrons. All four scattering lengths ($a_{gg}$, $a^\pm_{eg}$, and $a_{ee}$) are, thus, expected to be independent of the nuclear spin because both $g$ and $e$ have total electronic angular momentum $J$ equal to zero, which results in the decoupling between nuclear and electronic degrees of freedom during the course of a collision.
The decoupling during a
 collision is a consequence of the fact that each of the four
 molecular electronic states that correlate with the $J=0$ separated
 atom pair has zero projection $\Omega$ of total electronic angular
 momentum on the molecular axis. The nuclear spins in this case
 can only couple very weakly to other molecular states, even if there is a
 molecular curve crossing.

 While the short-range potential energy structure
 for a molecule like Sr$_2$ is very complex for the excited states
 \cite{Boutassetta1996X,Czuchaj2003X},
we will now show that scattering length differences among different
 combinations of nuclear spin projections for the same $\Omega=0$ potential
 are expected to be very small. The scattering length $a$ can be computed as $a = \bar{a} [1- \tan(\Phi - \pi/8)]$, where $\bar{a}$ is the average scattering length governed by the asymptotic behavior of the potential and 
 $\Phi$ is the semiclassical phase computed at zero energy from the classical turning point $R_0$ to infinity: $\Phi = \int_{R_0}^\infty d R \sqrt{M [- V(R)]}/\hbar$, where $-V(R)$ is the (positive) depth of the interaction potential at separation $R$ and $M/2$ is the reduced mass \cite{Gribakin1993X}. Defining $R(t)$ as the classical trajectory from time $t = -\infty$ to time $t = \infty$ of a particle of mass $M/2$ at zero energy in the potential $V(R)$, we can rewrite the phase as $\Phi = - \int_{-\infty}^{\infty} dt V(R(t)) /\hbar$. The order of magnitude of the change $\delta \Phi$ in the phase associated with different nuclear spin projections can, thus, be estimated as $\delta \Phi \sim \Delta t \delta V/\hbar$, where $\Delta t$ is the total time in the short-range part of the collision and $\delta V$ is the typical energy difference associated with different nuclear spin projections during this time. Since $\delta V$ vanishes at $R \to \infty$, only  the short range molecular region contributes to the phase difference. Therefore, assuming $\delta \Phi \ll 1$, $a \sim \bar{a}$, and $|\cos(\Phi - \pi/8)| \sim 1$, the nuclear-spin-dependent variation $\delta a$ in the scattering length can be estimated as $\delta a/a \sim \delta \Phi \sim  \Delta t \delta V/\hbar$.

Turning to the actual numbers, $\Delta t$ can be estimated from the depth ($\sim 10^3 \textrm{cm}^{-1} h c$) and the range ($\sim 10$ Bohr radii) 
of the appropriate interatomic potential (see e.g.~\cite{Boutassetta1996X,Czuchaj2003X}) to be $\Delta t \approx 1$ ps. For $g$-$g$ collisions,  $\delta V/h$ can be estimated by the second-order formula $E_\textrm{hf}^2/(h E_\textrm{opt}) \sim 200$ Hz, where $E_\textrm{hf}/h \sim  300$MHz is the approximate value for the hyperfine splittings in ${}^3P_1$ in ${}^{87}$Sr and $E_\textrm{opt}/h \sim 400$ THz is the optical energy difference between ${}^1S_0$ and ${}^3P_1$ in ${}^{87}$Sr. This yields the following estimate for the dependence of $a_{gg}$ on the nuclear spin: $\delta a_{gg}/a_{gg} \sim \delta \Phi \sim 10^{-9}$. For $e$-$e$ and $e$-$g$ collisions, an analogous second-order formula would use  the fine structure splitting between ${}^3P_1$ and ${}^3P_0$ in ${}^{87}$Sr ($E_\textrm{f}/h \sim 6$ THz) instead of $E_\textrm{opt}$ to yield $\delta \Phi \sim 10^{-7}$. However, the latter estimate ($\delta \Phi \sim 10^{-7}$) is too optimistic since molecular states that are split by $E_\textrm{f}$ at large interatomic separations may come orders of magnitude closer at short range \cite{Wang1998X}. Therefore, a more realistic conservative estimate would use the first-order formula $\delta V \sim E_\textrm{hf}$ to yield $\delta a_{ee}/a_{ee} \sim \delta a^\pm_{eg}/a^\pm_{eg} \sim \delta \Phi \sim 10^{-3}$. It is important to note, however, that these are all only very rough estimates. For example, hyperfine coupling in a molecule will differ from the hyperfine coupling in separated atoms. In fact, since it is very difficult to predict $\delta a/a$ accurately, these values would need to be measured. To conclude this Section, we would like to emphasize that, as mentioned in the main text, if the small nuclear-spin dependence of $a_{ee}$ and $a_{eg}^\pm$ is not negligible for some applications, one can use two different ground state atomic species instead of a ground and an excited state of one species.

\subsection*{Likelihood of Lossy $e$-$e$ Collisions and Possible Solutions}

Collisions of two $e$ atoms are likely to be accompanied by large loss \cite{Traverso2009X}. This can occur if the molecular $0_g^+$ potential that correlates with the $e$-$e$ atoms undergoes an avoided crossing with a potential curve that correlates with a lower energy pair of separated atoms (see, for example, Ref.~\cite{Czuchaj2003X}).  Similar crossings that result in inelastic energy transfer collisions were examined for $^1$P$_1 + ^1$S$_0$ collisions of alkaline earth atoms in Ref.~\cite{Machholm2001X}.  The likelihood of a relatively high probability of an inelastic event during such a crossing with species such as Sr or Yb means that the imaginary part $b_{ee}$ of the scattering length is expected to be large.  However, just like $a_{ee}$, $b_{ee}$ can not be calculated accurately from the potentials but would need to be measured.

The possible effects of $b_{ee}$ on the four examples we discuss   [Eqs.\ (\ref{H10}-\ref{KLM}) and Eq.\ (\ref{H11})] are as follows. $H_{(p,0)}$ [Eq.\ (\ref{Hp0})] is, of course, not affected because it involves only $g$ atoms. In $H_{(1,1)}$ [Eq.\ (\ref{H11})] and $H_{KLM}$ [Eq.\ (\ref{KLM})], the $e$ lattice is assumed to be so deep that $J_e$ is negligible compared to $U_{ee}+V_{ex}$ and $U_{ee}$, respectively, or to the experimental timescale, thus, fully suppressing tunneling of $e$ atoms and occupation of one site by more than one $e$ atom. The presence of an imaginary part $b_{ee}$ of the $e$-$e$ scattering length will give an effective nonzero width to the state with more than one $e$ atom per site and can, therefore, only further suppress this tunneling by a Zeno-like effect \cite{Syassen2008X,Daley2008X,Daley2009X}.

Therefore, $H_{(1,0)}$ [Eq.\ (\ref{H10})] is the only example that can be affected by large $b_{ee}$. In order for $H_{(1,0)}$ to contain a nonnegligible term proportional to $J_e^2/U_{ee}$, the ratio $|b_{ee}/a_{ee}|$ would need to be very small \cite{Tiesinga2000X}. 
Several approaches to avoid the losses associated with $b_{ee}$ in $H_{(1,0)}$ are possible. First, the large variety of stable atoms with two valence electrons (which includes not only alkaline-earths, but also Zn, Cd, Hg, and Yb) may have coincidentally an isotope with small $|b_{ee}/a_{ee}|$, which is more likely for lighter atoms \cite{Machholm2001X}. Second, while obtaining a good optical Feshbach resonance \cite{Ciurylo2005X, Naidon2006X, Enomoto2008X,Zelevinsky2006X,Escobar2009X} to reduce $|b_{ee}/a_{ee}|$ might not be possible, it should be possible to use optical Feshbach resonances to enhance $b_{ee}$ and, thus, suppress \cite{Syassen2008X,Daley2008X,Daley2009X} the virtual occupation of one site by two $e$ atoms; $H_{(1,0)}$ would then have the same form as in Eq.\ (\ref{H10}), except with $U_{ee}$ effectively set to infinity. Notice that here we suggest to use optical Feshbach resonances to affect $e$-$e$ scattering, which is different from the typical application to $g$-$g$ scattering \cite{Ciurylo2005X, Naidon2006X, Enomoto2008X,Zelevinsky2006X,Escobar2009X}. Third, one can consider using a different ground state atom to represent state $|e\rangle$, which would set $V_{ex} = 0$ in $H_{(1,0)}$. Finally, one could simply use an $e$-lattice that is deep enough to make $J_e$ negligible, which would, however, lead to the loss of terms in $H_{(1,0)}$ that exchange the pseudospin between neighboring sites.

\subsection*{Brief Review of Young Diagrams}

\begin{figure}[t]
  \begin{center}
  \includegraphics[scale=0.7]{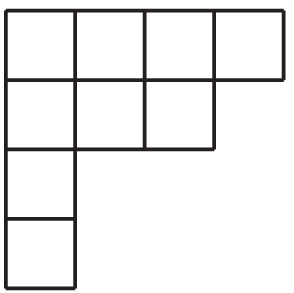}
  \end{center}
\begin{flushleft}
\noindent FIG. S1: \textbf{A general Young diagram.}
\end{flushleft}
\end{figure}

Irreducible representations of SU(2) are classified according to the total half-integer angular momentum $J$ and have dimension $2 J + 1$. On the other hand, a (semistandard) Young diagram, instead of a single value $J$, is used to describe an irreducible representation of SU(N) for a general $N$ \cite{Jones1998X, Fulton1997X}. As shown in the example in Fig.~S1, 
a Young diagram has all its rows left-aligned, has the length of rows weakly decreasing from top to bottom, and has at most $N$ rows. The dimension of the representation corresponding to a given diagram is the number of ways to fill the diagram with integers from $1$ to $N$ such that the numbers weakly increase across each row and strictly increase down each column. For our purposes, the number of boxes in the diagram is the number of atoms on the site, and the diagram describes the (nuclear) spin symmetry of the particular chosen single-site energy manifold. In particular, columns represent antisymmetrized indices, while rows are related to (but do not directly represent) symmetrized indices. It is the relation between antisymmetrized indices and the columns that limits the number of rows to $N$. On the other hand, since the full wavefunction (spin and orbital) on each site must satisfy complete fermionic antisymmetry, the relation between rows and symmetrized indices and the fact that we have only two orbital states ($g$ and $e$) force all our diagrams to have at most two columns. 

\subsection*{The $(p,q) = (1,1)$ spin Hamiltonian and the spin-1 Heisenberg antiferromagnet}

\begin{figure}[t]
  \begin{center}
  \includegraphics[scale=0.7]{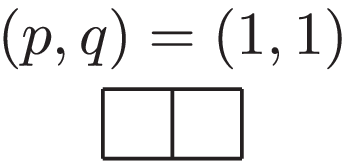}
  \end{center}
  \begin{flushleft}
\noindent FIG. S2: \textbf{(p,q) = (1,1) Young diagram.}
\end{flushleft}
\end{figure}

In the main text, we discussed two special cases of the spin Hamiltonian $H_{(p,q)}$, both of which had a single-column SU(N) representation on each site (i.e.~$q = 0$). In this Section, we discuss the simplest SU(N) representation with two columns, $(p,q) =
(1,1)$ [see Fig.~S2]. It  can be obtained when there is one $g$ and one $e$ atom
per site in the electronic singlet $|ge\rangle - |eg\rangle$
configuration. 
Setting $J_e = 0$ to avoid $e$-$e$ collisions, 
$H_{(p,q)}$ reduces 
to
 \begin{equation} \label{H11}
H_{(1,1)} =
\frac{J_g^2}{2 (U_{gg} + V_\textrm{ex})}
\sum_{\langle
i,j\rangle} S^2_{ij}. \tag{S1}
\end{equation}
The case of $N = 2$ is 
the spin-1 antiferromagnetic Heisenberg model. This model has a 1D ground state with
hidden topological structure 
 \cite{Girvin1989X}. 
 Recently, applications of related models in 
one-way quantum computation have been proposed \cite{Brennen2008X,Verstraete2004X}. 
Models with more complicated two-column representations may have exotic chiral spin liquid ground states that support non-Abelian anyons and that might thus be used for topological quantum computation \cite{Hermele2009X}.

\subsection*{The  Kugel-Khomskii model and the double-well phase diagram}

In the main text, we omitted the values of the parameters in $H_{(p,q)}$ that characterize the
Kugel-Khomskii model $H_{(1,0)}$  [Eq.\ (\ref{H10})]. In this Section, we present these parameters. We also present a detailed discussion of the double-well case phase diagram.

The parameters in $H_{(p,q)}$ that characterize the
Kugel-Khomskii model $H_{(1,0)}$  [Eq.\ (\ref{H10})]  are $\lambda_g^{ij} = - \kappa_g^{ij} =  \frac{2
J_g^2}{U_{gg}} \equiv -\kappa_g$, $\lambda_e^{ij} = -\kappa_e^{ij} =  \frac{2 J_e^2}{U_{ee}} \equiv -\kappa_e$,
$\kappa_{ge}^{ij} = - \frac{J_e^2+J_g^2}{2 U_{eg}^+} -
\frac{J_e^2+J_g^2}{2 U^-_{eg}} \equiv \kappa_{ge}$, $\lambda_{ge}^{ij} =
\frac{J_e^2+J_g^2}{2 U_{eg}^+} - \frac{J_e^2+J_g^2}{2 U^-_{eg}} \equiv \lambda_{ge}$,
$\tilde \kappa_{ge}^{ij} = \frac{J_e J_g}{U_{eg}^-} - \frac{J_e
J_g}{U_{eg}^+} \equiv \tilde \kappa_{ge}$, $\tilde \lambda_{ge}^{ij} = \frac{J_e J_g}{U_{eg}^-} +
\frac{J_e J_g}{U_{eg}^+} \equiv \tilde \lambda_{ge}$. To avoid loss in $e$-$e$ collisions, we assume for the rest of this Section that $U_{ee} = \infty$ 
(see Supplementary Information for a discussion of losses in $e$-$e$ collisions).

The nontrivial orbital-orbital, spin-spin, and spin-orbital interactions in $H_{(1,0)}$ [Eq.\ (\ref{H10})] result in competing orders, with the actual ground-state order dependent on the parameters of the Hamiltonian $H_{(1,0)}$. To get a sense of the possible orders, we consider the case $N = 2$ (with the spin states denoted by  $\uparrow$ and $\downarrow$) and discuss the double-well problem, with the wells denoted by $L$ (left) and $R$ (right). Due to the large optical energy separating $e$ and $g$, which we have ignored after Eq.~(\ref{ham0}), the three manifolds of constant $T^z = T^z_L + T^z_R$ ($T^z = -1, 0, 1$) should each be considered separately.

The four states in the $T^z = 1$ manifold, the subspace of two $e$ atoms, are $|ee\rangle |s\rangle$ and $|ee\rangle |t\rangle$. Here $|ee\rangle = |ee\rangle_{LR}$ is the orbital (or pseudo-spin) state, while $|t\rangle=|\uparrow \uparrow\rangle_{LR},|\downarrow \downarrow\rangle_{LR},\frac{1}{\sqrt{2}}(|\uparrow \downarrow\rangle_{L R} +|\downarrow \uparrow\rangle_{L R})$ and $|s\rangle=\frac{1}{\sqrt{2}}(|\uparrow\downarrow\rangle_{LR}-|\downarrow \uparrow\rangle_{LR})$ are the triplet and singlet spin states.  Since $U_{ee} = \infty$, all four of these states have zero energy and the ground-state phase diagram is trivial.

The four states in the $T^z = -1$ manifold (two $g$ atoms) are split by $H_{(1,0)}$ into two energy manifolds:
\begin{equation}
\!\!\!\!\!\!\!\!\!\!\!\!\!   |gg\rangle|t\rangle, \quad  E= 0,  \tag{S2}\\
\end{equation}
\vspace{-0.3in}
\begin{equation}
|gg\rangle|s\rangle, \quad E=-\frac{4 J_g^2}{U_{gg}}. \tag{S3}
\end{equation}
Only $|gg\rangle |s\rangle$ can take advantage of the virtual tunneling since two $g$ atoms in the triplet spin states cannot sit on the same site. Which of the two manifolds is the ground manifold depends on the sign of $U_{gg}$, as shown in the ground-state phase diagram in Fig.~\ref{KKfigure}a.  It is important to emphasize that for $U_{gg} < 0$, the subspace of one $g$ atom per site may be subject to extra loss down to the lower energy states that have both $g$ atoms in the same well. It is also worth noting that the diagram is only valid for $J_g \ll |U_{gg}|$.

Finally, the eight states in the $T^z = 0$ manifold (one $g$ atom and one $e$ atom) are split by $H_{(1,0)}$ into four energy manifolds:
\begin{equation}
|\Sigma\rangle|t\rangle, \quad  E=-\frac{(J_g+ J_e)^2}{U_{eg}^-}, \label{kksupp1} \tag{S4}
\end{equation}
\vspace{-0.2in}
\begin{equation}
|\tau\rangle|s\rangle, \quad E=-\frac{(J_g+ J_e)^2}{U_{eg}^+},  \tag{S5}
\end{equation}
\vspace{-0.2in}
\begin{equation}
|\tau\rangle|t\rangle, \quad  E=-\frac{(J_g- J_e)^2}{U_{eg}^-},  \tag{S6} 
\end{equation}
\vspace{-0.2in}
\begin{equation}
|\Sigma\rangle|s\rangle, \quad  E=-\frac{(J_g- J_e)^2}{U_{eg}^+}, \label{kksupp2} \tag{S7}
\end{equation}
where  $|\Sigma\rangle=\frac{1}{\sqrt{2}}(|e g\rangle_{LR}-|g e\rangle_{L R})$ and  $|\tau\rangle=\frac{1}{\sqrt{2}}(|e g\rangle_{L R}+|g e\rangle_{L R})$ are  anti-symmetric and symmetric orbital states, respectively. The denominators $U_{eg}^-$ and $U_{eg}^+$ in the  energies of the $|t\rangle$  and $|s\rangle$ states, respectively, reflect the fact that tunneling preserves the nuclear spin. At the same time, the $\pm$ signs in
the numerators can be understood by considering the case $J_g=J_e$, when all states with overall symmetry under particle exchange must have zero energy since for these states tunneling is forbidden due to the Pauli exclusion principle. The corresponding ground-state phase diagram as a function of the signs and relative magnitude of $U_{eg}^+$ and $U_{eg}^-$ is shown in Fig.~\ref{KKfigure}b.  As in the case of the $T^z = 1$ phase diagram, negative interaction energies may lead to increased losses.

\subsection*{Effects of Three-Body Recombination}

Three-body recombination \cite{Esry1999X, Bedaque2000X, Jack2003X, Kraemer2006X, Daley2009X} is a process during which three atoms come together to form a diatomic bound state and a single atom, and both final products have enough kinetic energy to leave the trap. While in certain cases, three-body recombination can be an asset \cite{Daley2009X}, usually it results in the loss of atoms and, thus, limits the duration of the experiment.  
 For our purposes, we can describe three-body recombination by a decay rate $\gamma_3$ \cite{Daley2009X} resulting in a loss of three particles from one site. This rate will likely depend on what atomic states are involved and, to the best of our knowledge, has not yet been measured or calculated for fermionic alkaline-earth atoms. 

Out of the four examples  [Eqs.\ (\ref{H10}-\ref{KLM}) and Eq.\ (\ref{H11})] that we discuss, only $H_{(1,1)}$ [Eq.\ (\ref{H11})] and $H_{(p,0)}$ [Eq.\ (\ref{Hp0})] may be affected by three-body recombination ($H_{KLM}$ [Eq.\ (\ref{KLM})] assumes negligible $g$-$g$ interactions, such as in a very shallow $g$ lattice or with a low density of $g$ atoms). In the case of $H_{(1,1)}$, two $g$ atoms and one $e$ atom occupy the same site virtually in the intermediate state that gives rise to the second order spin Hamiltonian with interaction strength $\propto J_g^2/(U_{gg} + V_{ex})$. Thinking of $\gamma_3$ as an effective linewidth for the intermediate state, $H_{(1,1)}$ will be valid and losses small provided that $\gamma_3$ is smaller than the effective "detuning" $U_{gg} + V_{ex}$. Since scattering lengths for alkaline-earth atoms \cite{Enomoto2008X, Escobar2008X,Campbell2009X}  are comparable to those for alkali atoms, $U_{gg} + V_{ex}$ can be on the order of several kHz \cite{Trotzky2008X}. At the same time, $1/\gamma_3$ for bosonic alkali atoms in deep traps can be on the order of 1 s \cite{Campbell2006X}. If $\gamma_3$ were the same in our case, $\gamma_3 \ll U_{gg} + V_{ex}$ would be satisfied. Ways of controlling the interactions via optical Feshbach resonances \cite{Ciurylo2005X, Naidon2006X, Enomoto2008X,Zelevinsky2006X,Escobar2009X} may also be envisioned.

In the case of $H_{(p,0)}$ [Eq.\ (\ref{Hp0})], $(n_A, n_B) = (1,1)$ does not suffer from three-body recombination. $(n_A, n_B) = (1,2)$ and $(2,2)$ may have three atoms per site virtually. As in the discussion of $H_{(1,1)}$, provided $\gamma_3$ associated with three $g$ atoms per site is smaller than $U_{gg}$, these configurations should be accessible. For the case $(n_A, n_B) = (1,2)$, $\gamma_3 \gg U_{gg}$ is also acceptable, since it will effectively prohibit the tunneling of the atoms to the state with 3 atoms on a site \cite{Daley2009X}, but the interaction can still take place through the intermediate state, in which an atom from a $B$ site tunnels to an $A$ site and back. One can also envision ways to use optical Feshbach resonance techniques \cite{Ciurylo2005X, Naidon2006X, Enomoto2008X} to induce large $\gamma_3$.
To be able to resolve the superexchange coupling  $\sim J_g^2/U_{gg}$ in cases where $n_A$ or $n_B$ is equal to 3, one must  have $\gamma_3 < J_g^2/U_{gg}$. Given that superexchange coupling can be as high as 1 kHz \cite{Trotzky2008X}, this condition should also be achievable. Although $n_A$ or $n_B$ greater than 3 will result in even shorter lifetimes \cite{Jack2003X}, there is a good chance that relatively large $n_A$ and $n_B$ can be achieved: at least, for bosonic alkali atoms in an $n=5$ Mott insulator state, the lifetime can still be as long as $0.2$ s  \cite{Campbell2006X}.

\subsection*{The $(p,0)$ spin Hamiltonian with $n_A = n_B \neq N/2$}

\begin{figure}[t]
  \begin{center}
   \includegraphics[scale = 1]{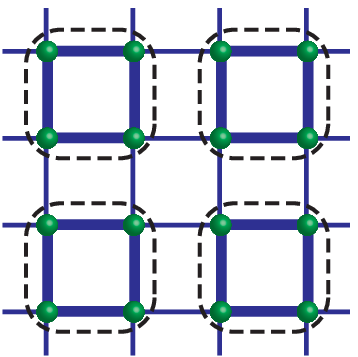}
  \end{center}
    \begin{flushleft}
\noindent FIG. S3: \textbf{Square lattice valence plaquette solid for $N=4$.} When $N = 4$ and $n_A = n_B = 1$, four sites are required to form an SU(4) singlet; these singlets can in turn form the schematically shown plaquette-ordered state  or a  disordered phase made of resonant plaquette states \cite{Bossche2000X}.
\end{flushleft}
\end{figure}

In the main text, we focused on one special case of the antiferromagnetic $(p,0)$ spin Hamiltonian on a square lattice, that with $n_A + n_B = N$ (where $n_A$ and $n_B$ denote the number of atoms per site on the two sublattices). In this Section, we describe another interesting and experimentally relevant case, $n_A = n_B \neq N/2$ 
\cite{Pankov2007X,Xu2008X,WangX,Bossche2000X,Greiter2007X,Rachel2009X,Sutherland1975X,Hermele2009X}.
Potential ground states include states built from valence \emph{plaquettes} (Fig.~S3)
\cite{Pankov2007X, Xu2008X}, resonant \emph{plaquette} states \cite{Bossche2000X},  and topological spin liquids
\cite{WangX,Hermele2009X}.  Valence plaquette states and resonant plaquette states are the natural generalization of
VBS states and resonant valence bond states (RVB) \cite{Anderson1987X}, respectively; for example, when $n_A = n_B = 1$, $N$
lattice sites are needed to form a SU(N) singlet. Fig.~S3 depicts a square lattice valence plaquette solid for $n_A = n_B = 1$ and $N=4$.
Techniques for detecting some of these phases are discussed in Ref.~\cite{Hermele2009X}. 
The experiment described in the main text for the case $n_A + n_B = N$ can also be generalized to probe the $n_A = n_B \neq N/2$ phase diagram including exotic phases such as valence plaquette solids [Fig.~S3], as well as competing magnetically ordered states. The main difference is that after preparing a band insulator of $N$ $g$ atoms per site, each site should be split not necessarily into two sites but into the number of sites that is appropriate for the case being considered (e.g.\ 4 for the case shown in Fig.~S3).


\subsection*{Double-well Kugel-Khomskii and RKKY experiments}

In the Methods, we have omitted  the description of how to prepare the initial states for the proof-of-principle double-well Kugel-Khomskii and RKKY experiments. We present this description in this Section.

\begin{figure}[t]
  \begin{center}
   \includegraphics[scale = 1]{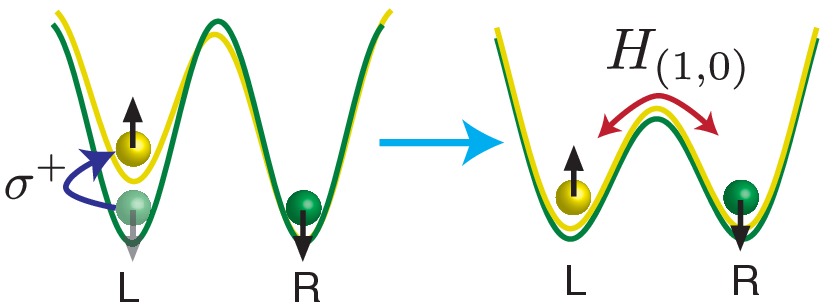}
  \end{center}
    \begin{flushleft}
\noindent FIG. S4: \textbf{A schematic diagram describing the preparation of the double-well state $|e,\uparrow\rangle_L|g,\downarrow\rangle_R$.}
\end{flushleft}
\end{figure}

We first describe how to prepare an array of independent double wells in the state $|e,\uparrow\rangle_L|g,\downarrow\rangle_R$, which we use for the proof-of-principle experiment to probe the spin-orbital interactions in the Kugel-Khomskii model \cite{Arovas1995X,Li1998X,Pati1998X, Zasinas2001X,Li2005X,Itoi2000X, Zhang2001X}. After loading a band insulator of $|g,\downarrow\rangle$ atoms in a deep optical lattice, an additional lattice for both $g$ (green) and $e$ (yellow) atoms with twice the spacing of the first lattice is turned on in one direction to create an array of independent double wells  \cite{Trotzky2008X}. Then, as shown in Fig.~S4, in the presence of an $e$-lattice bias, $\sigma^+$ polarized light on resonance with the $|g,\downarrow\rangle_L\to |e,\uparrow\rangle_L$ transition can be used to prepare the state $|e,\uparrow\rangle_L|g,\downarrow\rangle_R$. For examples of earlier orbital physics studies with ultracold atoms, where the orbitals are distinguished only by the different motional states of the atoms, we refer the reader to Refs.~\cite{Wu2008X, Zhao2008X, Xu2007X, Wang2008X, Muller2007X, Anderlini2007X} and references therein.


\begin{figure}[t]
  \begin{center}
  \includegraphics[scale=1]{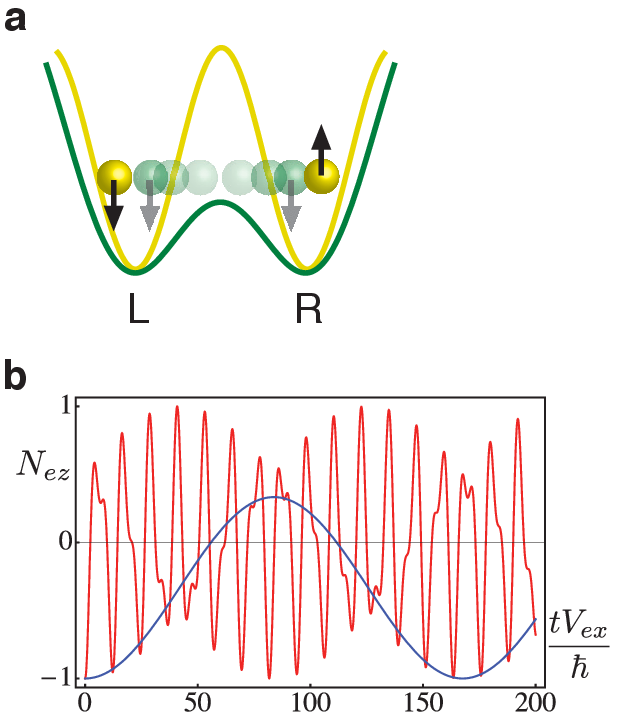}
  \end{center}
\begin{flushleft}
\noindent FIG. S5: \textbf{Proof-of-principle experiment to probe RKKY interactions in an array of isolated double wells.} \textbf{a,}  Schematic representation of the initial state $\frac{1}{\sqrt{2}}(|g,\downarrow\rangle_L+|g,\downarrow\rangle_R)|e,\downarrow\rangle_L|e,\uparrow\rangle_R$.
\textbf{b,} In the limit $|V_{ex}| \ll J_g$, the Neel order parameter for the $e$ atoms [$N_{ez}(t) = \frac{1}{2}[n_{e\uparrow L}-n_{e\downarrow L}-(n_{e\uparrow R}-n_{e\downarrow R})]$] is 
$N_{ez}(t) \approx -\frac{1}{3} \cos \left(\frac{V_{ex} t}{\hbar}\right) - \frac{2}{3} \cos \left(\frac{V_{ex} t}{2 \hbar} - \frac{3 V_{ex}^2 t}{8 J_g \hbar}\right)$, which is shown in red for $V_{ex} = - J_g/10$. It exhibits fast oscillations with frequency $\sim V_{ex}$, modulated by an envelope of frequency $\sim V_{ex}^2/J_g$ induced by RKKY interactions ($-\frac{1}{3} - \frac{2}{3} \cos \left(\frac{3 V_{ex}^2 t}{8 J_g \hbar}\right)$  shown in blue).
\end{flushleft}
\end{figure}

We now describe how to prepare the initial state $\frac{1}{\sqrt{2}}(|g,\downarrow\rangle_L+|g,\downarrow\rangle_R)|e,\downarrow\rangle_L|e,\uparrow\rangle_R
$ (see Fig.~S5a) for the double-well proof-of-principle RKKY experiment, whose expected Neel order parameter $N_{ez}(t)$ for $V_{ex} = - J_g/10$ is show in Fig.~S5b.
The first step to prepare the initial state
$\frac{1}{\sqrt{2}}(|g,\downarrow\rangle_L+|g,\downarrow\rangle_R)|e,\downarrow\rangle_L|e,\uparrow\rangle_R
$ is to load a
band insulator with three $|g,\downarrow\rangle$ atoms per site on
the long lattice and then slowly ramp  up the short lattice with a
bias so that it is energetically favorable to have two atoms in
the left well and one in the right well. Next one can change the
state of the right atom from  $|g,\downarrow\rangle_R$ to
$|e,\uparrow\rangle_R$ by applying a $\pi$ pulse of $\sigma^+$
polarized light  resonant with this single-atom transition. The
left well will be unaffected because the spectrum is modified by
the interactions (if interactions alone do not provide the desired
selectivity, one could, for example, change the bias of the
$e$-lattice). The next step is to change the state of the left
well from two $|g,\downarrow\rangle_L$ atoms populating the lowest
two vibrational states to $|e,\downarrow\rangle_L
|g,\downarrow\rangle_L$ both populating the lowest vibrational
state. This can be accomplished by using $\pi$-polarized traveling
wave laser light to apply a $\pi$  pulse resonant with the
transition between these two many-body states  \cite{Muller2007X}.
This results in
$|e,\downarrow\rangle_L|g,\downarrow\rangle_L|e,\uparrow\rangle_R$.
One can then temporarily shift  the $g$ and $e$ lattices relative to each
other to set $U^\pm_{eg}$ interactions to zero, then make $J_g$
nonzero, and wait until  the $g$ atom evolves  into  the desired
superposition
$\frac{1}{\sqrt{2}}(|g,\downarrow\rangle_L+|g,\downarrow\rangle_R)$
via tunneling. This yields the desired state
$\frac{1}{\sqrt{2}}(|g,\downarrow\rangle_L+|g,\downarrow\rangle_R)|e,\downarrow\rangle_L|e,\uparrow\rangle_R
$.




\end{document}